\documentclass[showpacs,superscriptaddress,amsmath,aps]{revtex4}
\usepackage{graphicx,color}
\usepackage{CJK}
\usepackage{bm}
\usepackage[hypertex]{hyperref}

\newcommand{\be}{\begin{equation}}
\newcommand{\ee}{\end{equation}}
\newcommand{\bea}{\begin{eqnarray}}
\newcommand{\eea}{\end{eqnarray}}
\newcommand{\bsube}{\begin{subequations}}
\newcommand{\esube}{\end{subequations}}

\newcommand{\Eq}[1]{Eq.\,(\ref{#1})}

\newcommand{\dg}{\dagger}
\newcommand{\la}{\langle}
\newcommand{\ra}{\rangle}

\newcommand{\beq}{\begin{equation}}
\newcommand{\eeq}{\end{equation}}
\newcommand{\beqn}{\begin{eqnarray}}
\newcommand{\eeqn}{\end{eqnarray}}
\newcommand{\bsub}{\begin{subequations}}
\newcommand{\esub}{\end{subequations}}

\begin{document}
\begin{CJK*}{GBK}{Song}

\title{Counting statistics of photon emissions
detected in non-Markovian environment }

\author{Luting Xu}
\email{xuluting@tju.edu.cn}
\affiliation{Center for Joint Quantum Studies and Department of Physics, Tianjin University,
Tianjin 300072, China}
\affiliation{ Department of Physics, Beijing Normal University,
Beijing 100875, China}

\author{Xin-Qi Li}
\email{xinqi.li@tju.edu.cn}
\affiliation{Center for Joint Quantum Studies and Department of Physics, Tianjin University,
Tianjin 300072, China}
\affiliation{ Department of Physics, Beijing Normal University,
Beijing 100875, China}

\date{\today}

\begin{abstract}
{\flushleft In this work}
we present a large-deviation analysis
for the counting statistics of atomic spontaneous emissions
continuously detected in finite-bandwidth non-Markovian environment.
We show that the statistics of the spontaneous emissions
depends on the time interval ($\tau$) of successive detections,
which can result in big differences
such as dynamical phase transition.
This feature excludes the idea
of regarding the spontaneous emissions
as detection-free {\it objective} events.
Possible experiment is briefly discussed
in connection with the state-of-the-art optical cavity set-up.
\end{abstract}

\maketitle

{\flushleft In quantum theory }
the problem that the spontaneous emission of photon from an atom
is dynamically {\it objective}
or {\it detector-dependent}
is fundamentally important and interesting.
At the early stage the quantum `jumps'
associated with the photon emissions were conceived of
as objective dynamical events \cite{Boh13,Ein17}.
However, the later development of quantum mechanics
within the framework of quantum wavefunction description
implies that the photon emissions can take place only
by detection (measurement) \cite{Dali92,WM93,WM09,Jac14},
in marked contrast to the objective jumps of Bohr and Einstein.
Very recently, this problem
was revisited by Wiseman {\it et al.}
by showing how different detection schemes
can result in different types of jumps \cite{Wis12},
in terms of quantum-mechanically steering
the `earlier emission event' by the post-stage detection.

In this work we alternatively make this issue in contact with
the counting statistics of the spontaneous emissions.
To exclude the picture as objective events,
we show that the spontaneous emissions
are strongly affected by
the time interval ($\tau$) in between
the moments we check the emissions happened or not.
We also show that this demonstration can be fulfilled
only by performing the photon detections in
a finite-band non-Markovian reservoir \cite{Zhang17,Zhang10}.
Associated with the non-Markovian dynamics of open quantum
systems \cite{Zhang17,Zhang10}, existing stochastic unraveling evolution
of the reduced density matrix dynamics
cannot be interpreted as measurement-conditioned physical quantum trajectory
\cite{Str98,Str99,Wis08,Dio08,Wis03}.
This is in sharp contrast with the situation of photon-detections
in (infinite) wide-band Markovian environment \cite{Dali92,WM93,WM09,Jac14},
where the {\it no-effect} of intermediate frequent
null-result (no emission registered) measurements
makes the ensemble average of the quantum trajectories
identical to the usual reduced density matrix.
We may explain the {\it no-effect} issue in more detail
by taking the simple example of an atom
subject to no more driving but prepared
in a superposition of the ground and excited states.
Let us imagine to check a photon emitted or not in the environment,
over the time duration $(0,t)$.
In the Markovian case, the many-times of frequent check over $(0,t)$
will arrive to the same conclusion
as that checking only at the last moment $t$.

\begin{figure}
\includegraphics[scale=0.5]{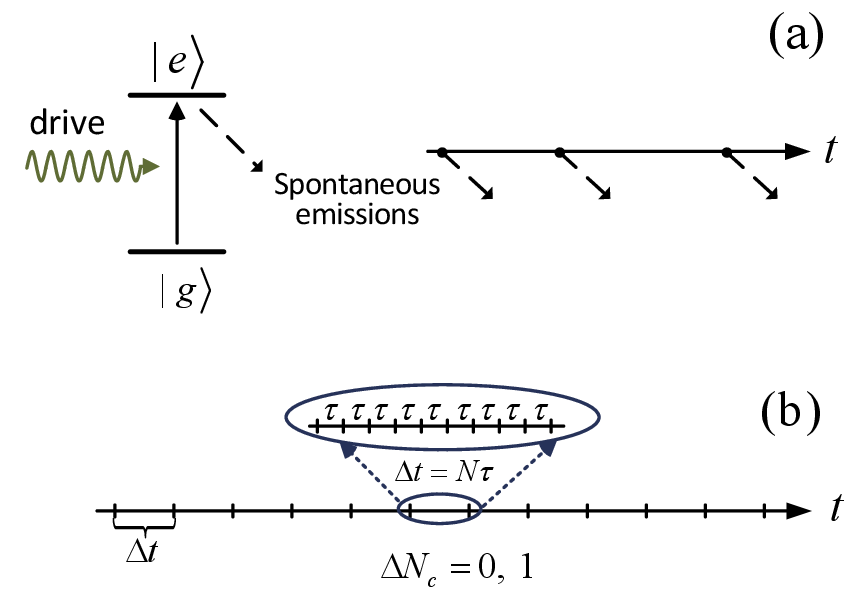}
\caption{
(a)
Schematic plot for the random spontaneous
photon emissions from a driven atom.
In the absence of photon detection
by introducing outside detector,
whether or not the spontaneous emissions take place
as {\it objective} events is of essential importance
which actually touches the bottom of quantum theory.
(b)
Successive detection of photons
after every short time interval $\tau$
(to mimic the `continuous' detection
by detector with response time $\tau$).
In order to construct an efficient theory for this type of
measurements, the accumulated result over $\Delta t=N\tau$
(determined by $\Delta N_c=0$ or 1), can be utilized
to perform a one-step update for the atom state.   }
\end{figure}

For atoms subject to continuous driving (Rabi oscillation),
as schematically shown in Figure.\ 1,
a series of spontaneous photon emissions will take place.
We can thus insert the above consideration into the study
of counting statistics of the spontaneous emissions,
in particular performing a large-deviation (LD) analysis
\cite{LD1,LD2,LD3,Gar10,Li11}.
We will show that the results would strongly
depend on the time interval $\tau$
in between the successive photo-detections, which can
lead to big differences such as dynamical phase transition.
In practice, the time interval $\tau$ in this theoretical consideration
qualitatively corresponds to the response time of photo-detectors.
Here the `response time' means the time delay of the output photo-current
after the photon to be measured reaches the detector.
This is the minimal time interval which allows us
to be able to count two successive photons.
Note also that, in the standard continuous-photon-detection-based
quantum trajectory theory (associated with measurements
in Markovian environment), this type of consideration
has been involved as well in constructing
the continuous measurement theory \cite{Dali92,WM93,WM09,Jac14}.   \\
\\
{\bf\large Results}\\
{\bf Model and measurement-results conditioned evolution.} \hspace{0.1cm}
Let us consider a driven multi-level atom coupled
to the electromagnetic vacuum (environment).
For the sake of simplicity,
we assume only a single radiative channel,
e.g., from $|e_j\ra$ to $|g\ra$.
The total Hamiltonian can be formally expressed as
\begin{align}\label{model}
H= H_S + \sum_k\left(b^{\dg}_{k}b_k+ 1/2 \right)\omega_k
+\sum_{k} \left[V_{k} b^{\dg}_{k}\sigma^-_j
     + {\rm H.c.} \right] \, .
\end{align}
Throughout this work we set $\hbar=1$.
The optical driving is contained in $H_S$,
and the coupling to the environment is via
the atomic operators $\sigma^-_j=|g\ra \la e_j|$
and $\sigma_j^+=|e_j\ra \la g|$. $V_{k}$ is the
coupling amplitude of the atom with the environment.
The property of the environment (and of the coupling) is largely
characterized by the spectral density function (SDF)
\bea
D(\omega)=\sum_{k}|V_{k}|^2
\delta(\omega-\omega_{k})
\to D_0\Lambda^2/[ (\omega-\omega_0)^2+\Lambda^2] \,.
\eea
Here we approximated the SDF by a finite-band Lorentzian
spectrum with $\omega_0$ the spectral center
and $\Lambda$ the width.

In the absence of detection,
the state of the whole system-plus-environment
evolves following the Schr\"odinger equation,
under the Hamiltonian of \Eq{model}.
However, the presence of detection in the reservoir
would interrupt this unitary evolution,
resulting in the `event' of photon emission (quantum `jump').
Conceptually, we assume that one is able to perform this
instantaneous detection after every short time interval $\tau$.
This is equivalent to the {\it continuous detection} by using
real detectors with signal-response time $\tau$ (see Figure.\ 1).

To construct an {\it efficient} theory for the successive
photon detections with very short time interval $\tau$
(to mimic the `continuous' detection),
one can utilize the accumulated result over $\Delta t=N\tau$
to perform a one-step update for the atom state, see Figure.\ 1(b).
This longer time duration $\Delta t$
is determined from the assumption that during $\Delta t$
there is {\it at most} one photon registered in the detector
\cite{Dali92,WM93,WM09,Jac14}.
Specifically, let us consider the time interval $(t,t+\Delta t)$.
There will be two possible outcomes:
a photon registered in the detector ($\Delta N_c=1$),
or no photon registered ($\Delta N_c=0$).
In the former case, we simply update the atom state
by a `jump' action; while for the latter result the atom
takes an {\it effective} smooth (but non-unitary) evolution.
Including also the evolution caused by the optical driving,
we can update the atom state in a compact way as
\bea\label{M01}
|\Psi(t+\Delta t)\ra = {\cal U}(\Delta t)
{\cal M}_{1,0}(\Delta t) |\Psi(t)\ra
\, /\parallel\bullet \parallel  \,,
\eea
where $\parallel\bullet \parallel$ denotes the normalization factor.
${\cal U}(\Delta t)$ describes the unitary evolution
owing to the optical driving, while ${\cal M}_{1,0}(\Delta t)$
are the Krause operators in the POVM formalism which read, respectively,
${\cal M}_{1}(\Delta t)=\sigma^-_j$ for $\Delta N_c=1$,
and ${\cal M}_{0}(\Delta t)=diag\{\bar{a}(\Delta t),1,\cdots,1 \}$
for $\Delta N_c=0$ and with $\bar{a}(\Delta t)$
given by \cite{SG14,Xu16}
\begin{align}
\bar a(\Delta t)=\exp\left\{
-\left[\frac{1}{c}-( 1-e^{-c x})
\frac{1}{c^2x}
\right]\frac{\Gamma \Delta t}{2}  \right\}   \,.
\label{scal}
\end{align}
In this elegant result, we have introduced
the `usual' emission rate $\Gamma=2\pi D_0$,
the frequency off-set parameter via
$E=(E_j-E_g)-\omega_0=d\, \Lambda$ and $c=1-id$,
and the scaling variable $x=\Lambda \tau$.
Note also that the above form of ${\cal M}_{0}(\Delta t)$
is associated with expressing the atom state
$|\Psi(t)\ra  = \alpha_j(t)|e_j\ra + \beta(t)|g\ra + \cdots $
in terms of a column vector
$[\alpha_j(t),\beta(t),\cdots ]^T$,
which makes well defined the action of
${\cal M}_{0}(\Delta t)$ on the atom state.

From \Eq{scal}, in the wide-band (Markovian) limit,
$x\to\infty$ and $c\to 1$, one recovers the standard result
$\bar a(\Delta t)\to e^{-\Gamma \Delta/2}$.
On the other hand, in the limit of $x\to 0$,
one finds from \Eq{scal} that $\bar a(\Delta t)=1$,
so that the atom is frozen in its initial state
under frequent measurements, showing the Zeno effect.
From \Eq{scal}, one can also define an {\it effective} decay rate
\bea\label{rate-1}
\gamma_{\rm eff}(x)={\rm Re}\left\{\left[ 1- (c x)^{-1}
\left( 1-e^{-c  x} \right) \right]/c \right\}\, \Gamma  \,.
\eea
Note that for the wide-band-limit Markovian environment,
the result implies {\it no-effect}
of the intermediate {\it null-result} (no photon detected)
interruptions \cite{SG13} .
For finite-bandwidth environment, however,
\Eq{rate-1} shows that the decay rate is influenced
by the frequent null-result measurements.
This $x$- or $\tau$-dependence is essentially rooted in
the non-Markovian nature of the environment. \\
\\
{\bf  Large-deviation analysis.} \hspace{0.05cm}
Below we outline the formalism for analyzing
the statistical properties of the
dynamical trajectories of the spontaneous emissions
\cite{LD1,LD2,LD3,Gar10,Li11}.
Actually, counting statistics of spontaneous emissions
is associated with the {\it ensemble average}
over the two possible outcomes leading to \Eq{M01}.
The resultant atom state is thus described by a reduced density
matrix which satisfies a master equation \cite{Xu16}.
For the purpose of large-deviation analysis,
we introduce the $n$-dependent
reduced density matrix, $\rho^{(n)}(t)$.
It describes the atom state conditioned on the total
number ($n$) of photons detected over $(0,t)$.
The equation-of-motion of $\rho^{(n)}(t)$ is given by \cite{Li11}
\bea\label{n-ME}
\dot{\rho}^{(n)}= -i[H_S,\rho^{(n)}]
+ \gamma_{\rm eff}(x)
\left( \sigma_j^- \rho^{(n-1)} \sigma_j^+ \right.  
- \left.\frac{1}{2} \{ \sigma_j^+\sigma_j^-,\rho^{(n)}\} \right) \;.
\eea
Knowing the $n$-resolved density matrix, we can obtain
the LD function $P(s,t)$ via the following transformation
\bea\label{LD-1}
P(s,t)=\sum_n e^{-s n} P(n,t)=e^{-{\cal F}(s,t)},
\eea
where $P(n,t)={\rm Tr}[\rho^{(n)}(t)]$
and, as to be clear below, ${\cal F}(s,t)$
plays the role of {\it generating} function for the LD analysis.
In \Eq{LD-1}, the real nature of the transforming factor $e^{-s n}$
makes the resultant $P(s,t)$ resemble the
partition function in statistical mechanics.
That is, the trajectories are categorized
by a dynamical order parameter ``$n$" or its conjugate field ``$s$".
In statistical mechanics, the partition function measures the number of
microscopic configurations accessible to the system under given conditions.
For the spontaneous emissions,
if we are interested in the dynamical aspects of the emitted photons,
the above insight can lead to an LD analysis in time domain.
In particular, it allows to inspect the {\it rare fluctuations}
or {\it extreme events} by tuning the conjugate field ``$s$".

In practice, instead of solving \Eq{n-ME},
we introduce $\rho(s,t)=\sum_n e^{-sn}\rho^{(n)}(t)$
to obtain the equation for $\rho(s,t)$ \cite{Li11}, and are able
to straightforwardly compute the LD function $P(s,t)$
by noting that $P(s,t)={\rm Tr}[\rho(s,t)]$.
Then, from the generating function
${\cal F}(s,t)=-\ln P(s,t)$, we have
\begin{subequations}\label{F-k}
\begin{align}
{\cal F}_1(s,t)&\equiv \partial_s {\cal F}(s,t)
=\frac{1}{P(s,t)}\sum_n n e^{-sn} P(n,t)
  \equiv \la n \ra_s \; ,   \\
{\cal F}_2(s,t)&\equiv \partial^2_s {\cal F}(s,t)
=-\la (n-\bar{n}_s )^2\ra_s  \;,
\end{align}
and more generally,
\begin{align}
{\cal F}_k(s,t)\equiv \partial^k_s {\cal F}(s,t)
=(-)^{(k+1)}\la (n-\bar{n}_s )^k\ra_s  \;.
\end{align}
\end{subequations}
Here, for brevity, we utilized also the notation
$\bar{n}_s$ for $\la n \ra_s$.
From these cumulants, we can define a finite-counting-time
flux of the emitted photons $I(s,t)={\cal F}_1(s,t)/t$
and the shot noise $S(s,t)=2 |{\cal F}_2(s,t)|/t$.
Conventionally, one may employ the Fano factor
$F(s,t)={\cal F}_2(s,t)/{\cal F}_1(s,t)$,
or the so-called Mandel factor
$Q(s,t)=-{\cal F}_2(s,t)/{\cal F}_1(s,t)-1$,
to characterize the fluctuation properties.

\begin{figure}
\centering
\includegraphics[scale=0.35]{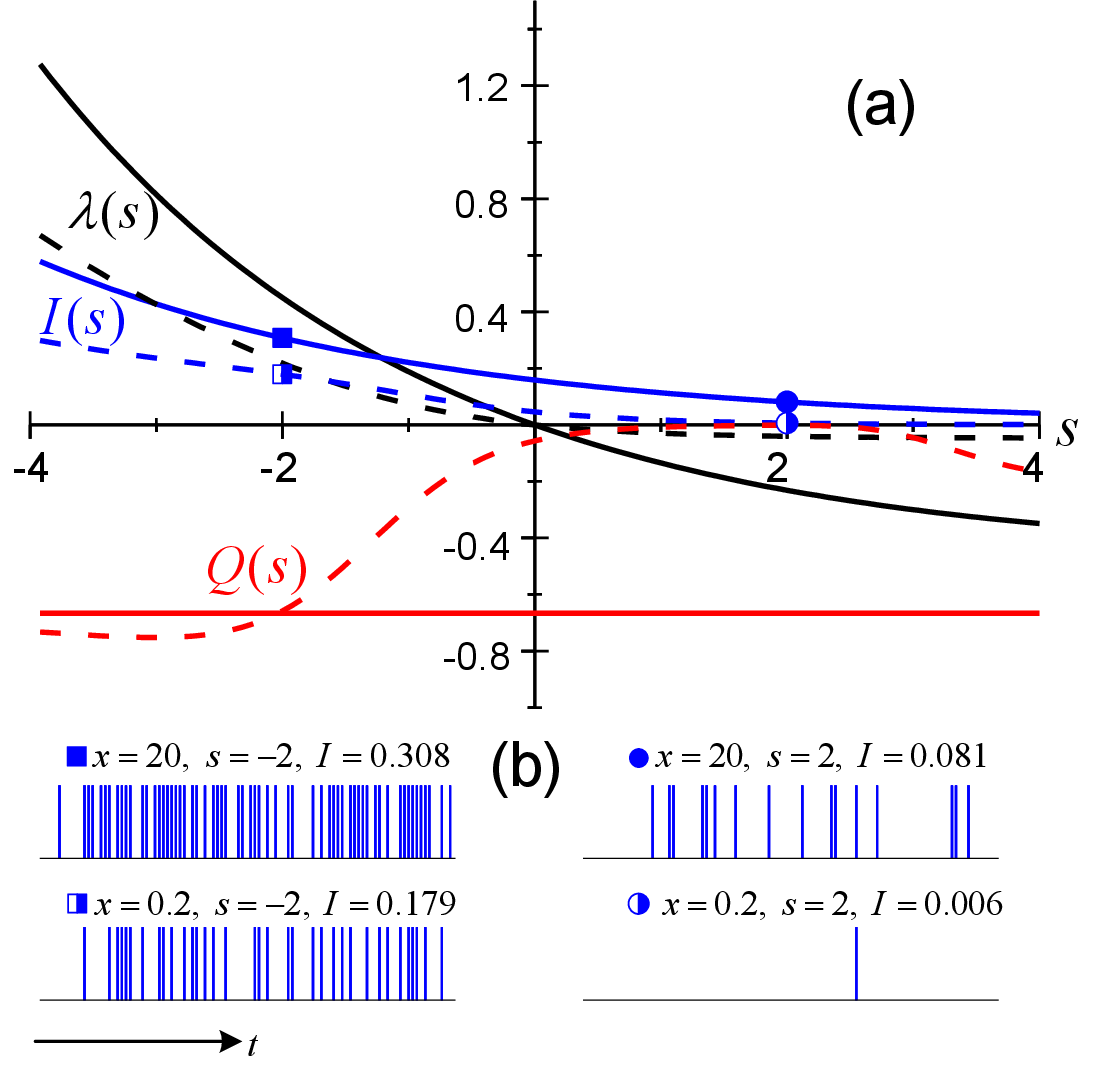}
\caption{ (Color online)
Large-deviation analysis for the spontaneous
emission trajectories of a driven two-level atom.
(a)
The characteristic function $\lambda(s)$,
the flux $I(s)$ of emitted photons,
and the Mandel factor $Q(s)$
are presented for two sets of trajectories
collected by photo-detectors with
different response times,
which correspond to the ``scaling" parameters
$x=20$ (solid lines) and $0.2$ (dashed lines).
We use a reduced units of system by setting the
``natural" spontaneous emission rate $\Gamma=1$,
and $\gamma_{\rm eff}|_{x=20}=4\Omega$.
(b)
Representative trajectories from the
sub-ensembles as indicated by the specific parameters. }
\end{figure}

\vspace{0.1cm}
{\flushleft\bf Model (I): two-level atom.}  \hspace{0.1cm}
First we consider a driven two-level atom, described by the Hamiltonian
$H_S=\frac{\Delta}{2}\sigma_z+\Omega\sigma_x$,
where $\sigma_z=|e\ra \la e|-|g\ra \la g|$
and $\sigma_x=|e\ra \la g|+|g\ra \la e|$.
The damping operator (spontaneous emission from $|e\ra$ to $|g\ra$)
in \Eq{n-ME} is simply given by $\sigma^-=|g\ra \la e|$.

The simulation result is displayed in Figure.\ 2.
For better understanding to the result presented here,
we mention that the LD function around $s=0$
encodes information of the {\it typical} trajectories,
while away from $s=0$, on the other hand,
it encodes information about the {\it rare} trajectories
via assigning a weight factor $e^{-sn}$ to select mainly
the {\it active} trajectories (for $s<0$),
or the {\it inactive} ones (for $s>0$).
In this plot (and in Figure.\ 3 in the following),
we consider a long counting time limit.
In this case it can be proved \cite{Li11},
that the generating function has an asymptotic form
${\cal F}(s,t)\simeq t\lambda(s)$,
and call $\lambda(s)$ the LD {\it characteristic} function.
In Figure.\ 2, the LD characteristic function $\lambda(s)$,
the $s$-dependent flux $I(s)$ of the emitted photons,
and the fluctuations -- the Mandel factor $Q(s)$ --
are plotted {\it versus} the conjugated field $s$,
as a `multi-angle' characterization
for the photon emission trajectories.

The essential point we may stress here is that
the statistical properties of the emission trajectories
depend on {\it how often we perform the detections},
in the sense as illustrated in Figure.\ 1(b).
That is, the trajectories {\it continuously collected}
by photo-detector with different response time $\tau$
may have quite different statistical properties.
Note that this is very different from the photon detection in
Markovian environment, where the result is $\tau$ independent.
For instance, in Figure.\ 2 we see that
for two different response times,
which result in $x=20$ and $0.2$,
the photon emission flux $I(s)$ with $x=20$ (larger $\tau$)
is stronger than the result with $x=0.2$ (smaller $\tau$).
In particular, the flux $I(s=0)$ of the typical trajectories
in the case $x=0.2$ almost vanishes,
which actually indicates the Zeno effect
since the very frequent detections
prevent the spontaneous emission.

More interesting is the behavior of the fluctuations
of the $s$-dependent trajectories.
For $x=20$ (we have purposely chosen this parameter),
we see that the Mandel factor $Q(s)$ is an $s$-independent constant,
which means a homogeneous fluctuation property.
In other words, all the sub-ensemble trajectories
collected with $x=20$ have the same fluctuations.
However, if we alter the detection time interval ($x=0.2$),
the fluctuations of the sub-ensemble trajectories
are no longer homogeneous, but $s$-dependent
as shown in Figure.\ 2 by the $Q(s)$ curve.

\begin{figure}
\centering
\includegraphics[scale=0.35]{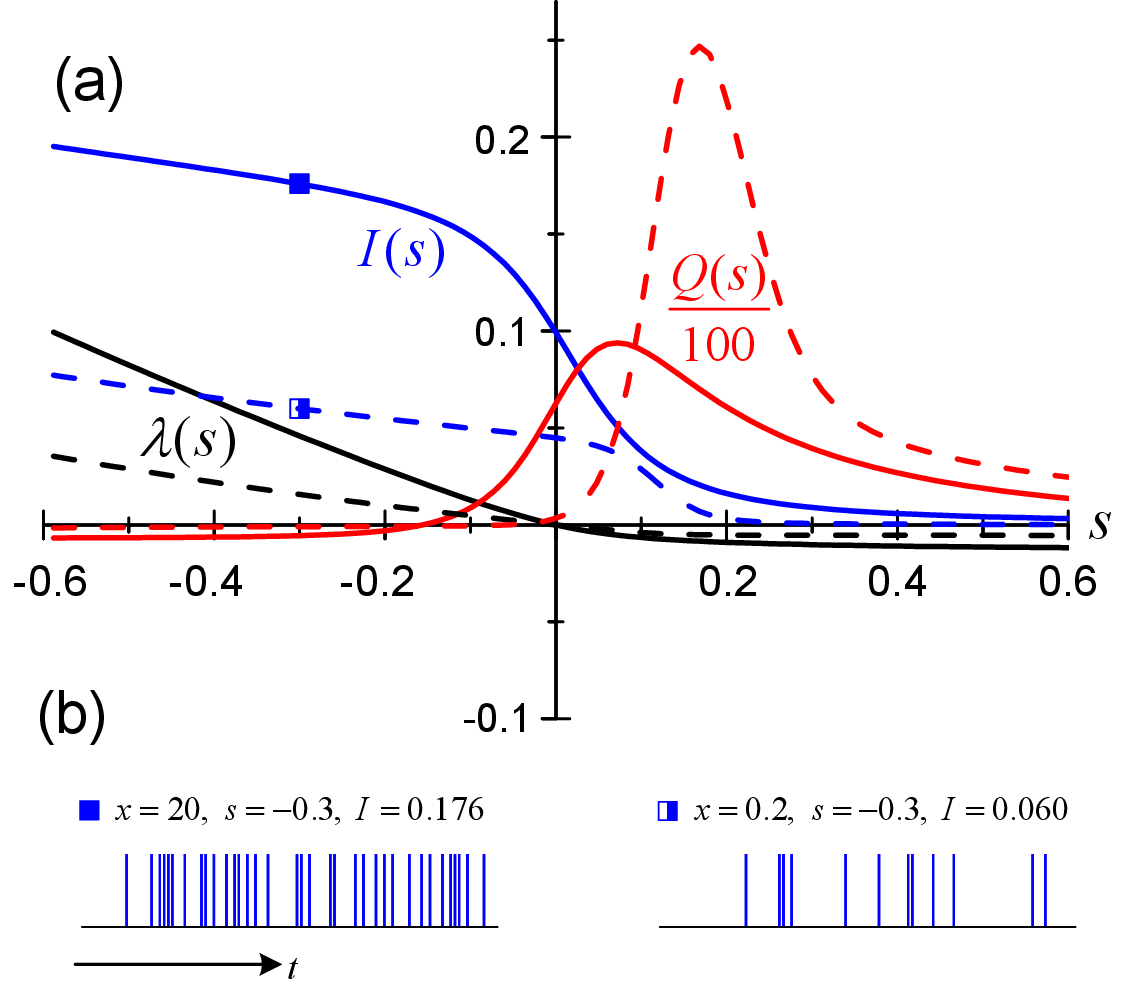}
\caption{ (Color online)
Large-deviation analysis for a driven three-level atom.
(a) and (b): The same plots as described in Figure.\ 2,
and similar reduced units of system adopted by setting
$\Gamma=1$, $\gamma_{\rm eff}|_{x=20}=4\Omega_1$,
and $\Omega_2=0.1\Omega_1$.  }
\end{figure}

\vspace{0.5cm}
{\flushleft\bf Model (II): three-level atom.}  \hspace{0.1cm}
The second example is the LD analysis for the spontaneous
emissions from a three-level atom.
The atom is driven by two resonant lasers
with Rabi couplings $\Omega_1$ and $\Omega_2$,
describe by the Hamiltonian
$H_S= \sum_{j=1,2}(\frac{\Delta_j}{2}\sigma_{jz}
+ \Omega_j \sigma_{jx})$,
where
$\sigma_{jz}=|e_{j}\ra\la g| - |e_{j}\ra\la g|$ and
$\sigma_{jx}=|e_{j}\ra\la g| +|g\ra\la e_j|$.
We assume only one spontaneous emission channel,
i.e., from $|e_1\ra$ to $|g\ra$.
So the damping operator in \Eq{n-ME} reads
$\sigma^-_1=|g\ra\la e_1|$.

The result of LD analysis for this driven three-level atom
is shown in Figure.\ 3.
For $s<0$ the active phase corresponds to plentiful photons emitted
and most occupation of the state $|e_1\ra$,
while for $s>0$ the inactive phase means that
the number of emitted photons is small
and the occupation is largely in the state $|e_2\ra$.
(Note that the spontaneous emission from $|e_2\ra$
to $|g\ra$ is forbidden as we have assumed).
Compared to the two-level atom studied above,
in the active side ($s<0$), the behaviors are similar.
However, in the inactive side ($s>0$), the difference is remarkable.
The most prominent feature is the appearance of a `crossover'
behavior between two distinct dynamical phases.
This is most clearly revealed by the Mandel factor $Q(s)$,
where the `sharp peak' indicates the `crossover' between
two distinct phases (on the two sides of the peak),
as we vary the LD parameter ($s$) through the peak region.

Actually, the crossover behavior is something of
a {\it smoothed first-order phase transition}.
We may understand this interpretation in more detail as follows.
The peak of $Q(s)$ in Figure.\ 3
simply means strong fluctuations of the sub-ensemble trajectories,
which are a consequence of fact that the sub-ensemble
is a {\it mixture} of two types of trajectories,
i.e., the relatively active and inactive ones
(on the two sides of the `peak').
In alternative words, the sub-ensemble
is a mixture of two distinct dynamical phases.
The active phase is that on the left side of the peak
and the inactive phase is the one on the right side.
We know that
{\it coexistence of two distinct phases} is the physical reason
of strong fluctuations, which resembles actually what happens
at the critical point (critical temperature)
of the first-order (thermal dynamic) phase transition.
Since the strong fluctuations appear in the proximity around the peak
(but not precisely at a unique critical point of $s$),
we may say that, when crossing the round peak, the system experiences
a `smoothed' first-order dynamical phase transition, more specifically,
a transition from photon-emission-active phase to inactive phase.

The crossover behavior (of suffering a dynamical phase transition)
is a consequence of the interplay
between the two channels of driving, i.e.,
$|g\ra \Leftrightarrow  |e_1\ra$ and $|g\ra \Leftrightarrow  |e_2\ra$,
and that only on $|e_1\ra$ the photon emission is allowed
while on $|e_2\ra$ it is forbidden.
Similar statistics behavior of dynamical trajectories
was found also in the transport through
a parallel double-dot system with Coulomb blockade \cite{Li11}
where the interplay of the Coulomb blockade and quantum interference
induces two effective transport channels, one is slow and another fast.

Again, in Figure.\ 3, we plot the results from two sets of trajectories
with different photon-detection time intervals, i.e., $x=20$ and $0.2$.
We find that the crossover behavior for the $x=0.2$
trajectories is more striking.
In the inactive ($s>0$) regime, the characteristic function
$\lambda(s)$ is more flat and the flux $I(s)$ of the emitted photons
vanishes more rapidly, meanwhile the $Q(s)$ peak is much higher
and shifts towards larger conjugate field $s$
(more inactive subensemble trajectories).
{\it
We stress that this $\tau$-dependent feature is unique only for
continuous detection of photons in a non-Markovian environment,
which does not happen for detection in Markovian environment. }   \\
\\
{\large\bf Discussion} \\
We have presented a counting statistics study
at the level of large-deviation analysis,
for atomic spontaneous emissions continuously detected in
a non-Markovian environment with finite-bandwidth ($\Lambda$).
We showed that the statistics behaviors
can be strongly influenced by the response time ($\tau$)
of the detector, via the elegant scaling variable $x=\Lambda \tau$.
The feature that the trajectories of the spontaneous emissions
depend on how often we perform the detections
definitely excludes the idea of regarding the spontaneous emissions
as detection-free {\it objective} events.
This is because the detection interval $\tau$ is small enough
compared to the average time
between the successive spontaneous emissions,
thus there are no photons missed in the counting collection.
If the spontaneous emissions were {\it objective},
the statistical properties must be independent of $\tau$.
Via the scaling variable ($x=\Lambda \tau$) analysis,
we also showed that it is impossible to demonstrate
in Markovian environment the effect
of the detection time $\tau$ on the counting statistics.

In this work we have restricted our analysis to Lorentzian spectrum.
However, the above conclusion is valid to arbitrary SDF of non-Markovian environment
such as the Ohmic, sub-Ohmic, and super-Ohmic baths.
Actually, we have recently generalized the measurement theory
and the associated quantum trajectory approach
to environment beyond the Lorentzian spectrum \cite{Xu17}.
For arbitrary SDF, we proved in general the existence of scaling property.
Despite that analytical result is not available in general case,
we developed reliable numerical scheme to simulate the quantum trajectories.

As possible implementation in experiment,
one may consider to put the atom in
the state-of-the-art optical cavity.
The cavity mode coupled to outside (Markovian) world
is a good finite-bandwidth non-Markovian environment,
and is well described by the Lorentzian spectral density function.
One can then perform detection for the photons
leaked from the cavity.
In this set-up, the bandwidth $\Lambda$ can be modulated
by the leaky rate of the cavity photon,
to alter the scaling variable $x=\Lambda \tau$.
This is equivalent to altering the detection time $\tau$.

We notice that the spontaneous emissions (resonance fluorescence)
from driven artificial atom in superconducting circuit-QED system
have been detected in recent experiments
\cite{Hua14,Hua16,Mur16,Mur15,Mur16a,Jor16}.
However, owing to that direct detection of single photons
at microwave frequencies is not yet available at present stage,
the quadratures of the microwave-photon-field
are measured in these experiments,
based on the homodyne or heterodyne detections.
Statistics analysis of this type of measurement records
is an interesting open question worth future exploration,
especially from the perspective of measurement in
non-Markovian environment as considered in present work.
As a final remark, we mention also the recent interests
in the {\it most-likely-paths} (MLP) among the huge number
of stochastic quantum trajectories under continuous monitoring
\cite{Jor16,Jor13,Jor15}.
In this context, it would be of interest to study
the statistics of the sub-ensemble of `rare events' (rare paths),
in similar sense of the LD studies.

\vspace{1cm}
{\flushleft\bf\large Acknowledgements}\\
This work was supported by the National Key Research
and Development Program of China under No. 2017YFA0303304, the National Natural Science Foundation of China (NNSFC) under Nos. 11675016 and 21421003, the Beijing Natural Science Foundation under No. 1164014, the China Postdoctoral Science Foundation funded project under No. 2016M591103,
and the Fundamental Research Funds for the Central Universities under No. 2015NT12. \\
\\
{\bf\large Author Contributions}\\
X.Q.L. initiated the idea and supervised the work; L.X. carried out the analytic derivations
and numerical calculations; X.Q.L. wrote the paper and both authors reviewed it. \\
\\
{\bf\large Additional Information} \\
Competing financial interests: The authors declare no competing financial interests.

\end{CJK*}

\vspace{0.2cm}




\begin{references}




\bibitem{Boh13} 
Bohr, N. On the constitution of atoms and molecules. {\it Phil. Mag.} {\bf 26}, 1-25 (1913).
\bibitem{Ein17}
Einstein, A. Zur quantentheorie der qtrahlung. {\it Physikalische Zeitschrift} {\bf 18,} 121 (1917).
\bibitem{Dali92}   
Dalibard, J., Castin, Y., \& Molmer, K.  Wave-function approach to dissipative processes in quantum optics.
{\it Phys. Rev. Lett.} {\bf 68,} 580 (1992).
\bibitem{WM93}
Wiseman, H. M. \& Milburn, G. J. Quantum theory of field-quadrature measurements
{\it Phys. Rev. A} {\bf 47,} 642 (1993).
\bibitem{WM09}  
Wiseman, H. M. \& Milburn, G. J. Quantum measurement
and control (Cambridge Univ. Press, Cambridge, 2009).
\bibitem{Jac14}
Jacobs, K. Quantum measurement theory and
its applications (Cambridge Univ. Press, Cambridge, 2014).

\bibitem{Wis12}
Wiseman, H. M. \& Gambetta, J. M. Are dynamical quantum jumps detector dependent?
{\it Phys. Rev. Lett.} {\bf 108,} 220402 (2012).


\bibitem{Zhang17}   
Zhang, J., Liu, Y.-X., Wu, R.-B., Jacobs, K. \& Nori, F.
Quantum feedback: theory, experiments, and applications. {\it Phys. Rep.} {\bf 679,} 1-60 (2017).
\bibitem{Zhang10}
Zhang, J., Wu, R.-B., Li, C.-W. \& Tarn., T.-J. Protecting coherence and entanglement by quantum feedback controls. {\it IEEE Trans. Automat.  Contr. }{\bf 55,} 619-633 (2010).
\bibitem{Str98}
Di\'osi, L., Gisin, N. \& Strunz, W. T. Non-Markovian quantum state diffusion.
{\it Phys. Rev. A }{\bf 58}, 1699 (1998).
\bibitem{Str99}
Strunz, W. T., Di\'osi, L. \& Gisin, N. Open System Dynamics with Non-Markovian Quantum Trajectories.
{\it Phys. Rev. Lett.} {\bf 82}, 1801 (1999).
\bibitem{Wis08}
Wiseman, H. M. \& Gambetta, J. M. Pure-State Quantum Trajectories for General Non-Markovian Systems Do Not Exist.
{\it Phys. Rev. Lett.} {\bf 101}, 140401 (2008).
\bibitem{Dio08}
Di\'osi, L. Non-Markovian Continuous Quantum Measurement of Retarded Observables.
{\it Phys. Rev. Lett.} {\bf 100}, 080401 (2008).
\bibitem{Wis03}
Gambetta, J. \& Wiseman, H. M. Interpretation of non-Markovian stochastic Schr\"{o}dinger equations as a hidden-variable theory.
{\it Phys. Rev. A.} {\bf 68}, 062104 (2003).



\bibitem{LD1}           
Eckmann, J. P. \& Ruelle, D. Ergodic theory of chaos and strange attractors.
{\it Rev. Mod. Phys.} {\bf 57,} 617 (1985).
\bibitem{LD2}
Gaspard, P. Chaos, scattering and statistical mechanics
(Cambridge University Press, Cambridge, 2015).
\bibitem{LD3}
Touchette, H. The large deviation approach to statistical mechanics.
{\it Phys. Rep.} {\bf 478}, 1 (2009).
\bibitem{Gar10}
Garrahan, J. P. \& Lesanovsky, I. Thermodynamics of a quantum jump trajectories.
{\it Phys. Rev. Lett.} {\bf 104}, 160601 (2010).
\bibitem{Li11}
Li, J., Liu, Y., Ping, J., Li, S. S., Li, X.-Q. \& Yan, Y. J. Large-deviation analysis for counting statistics in mesoscopic transport.
{\it Phys. Rev. B }{\bf 84}, 115319 (2011).


\bibitem{SG14}
Xu, L., Cao, Y., Li, X.-Q., Yan, Y. J., \& Gurvitz, S. Quantum transfer through a non-Markovian environment under frequent measurements and Zeno effect.
{\it Phys. Rev. A} {\bf 90,} 022108 (2014).
\bibitem{Xu16}
Xu, L. \& Li, X.-Q. Quantum trajectories under frequent measurements in a non-Markovian environment.
{\it Phys. Rev. A} {\bf 94,} 032130 (2016).
\bibitem{SG13}  
Ping, J., Ye, Y., Li, X.-Q., Yan, Y. J. \& Gurvitz, S.  Undetectable quantum transfer through a continuum.
{\it Phys. Lett. A} {\bf 377,} 676 (2013).


\bibitem{Xu17}
Xu, L. \&  Li, X.-Q.
Theory for frequent measurements of spontaneous emissions
in non-Markovian environment: beyond Lorentzian spectrum.
Preprint at {\it https://arxiv.org/abs/1707.02578} (2017)


\bibitem{Hua14}   
Campagne-Ibarcq, P., Bretheau, L., Flurin, E.,
Auffeves, A., Mallet, F. \& Huard, B. Observing interferences between past and fFuture quantum states in resonance fluorescence.
{\it Phys. Rev. Lett.} {\bf 112,} 180402 (2014).
\bibitem{Hua16}
Campagne-Ibarcq, P., Six, P., Bretheau, L., Sarlette, A.,
Mirrahimi, M., Rouchon, P. \& Huard, B.
Observing quantum state diffusion by heterodyne detection of fluorescence.
{\it Phys. Rev. X} {\bf 6,} 011002 (2016).
\bibitem{Mur16}   
Foroozani, N., Naghiloo, M., Tan, D., Molmer, K. \& Murch, K. W.
Correlations of the time dependent signal and the state of a continuously monitored quantum system.
{\it Phys. Rev. Lett.} {\bf 116,} 110401 (2016).
\bibitem{Mur15}
Tan, D., Weber, S. J., Siddiqi, I., M{\o}lmer, K. \& Murch, K. W. Prediction and retrodiction for a continuously monitored superconducting qubit.
{\it Phys. Rev. Lett.} {\bf 114,} 090403 (2015).
\bibitem{Mur16a}
Naghiloo, M., Tan, D., Harrington, P. M., Lewalle, P.,
Jordan, A. N. \& Murch, K. W. Quantum caustics in resonance fluorescence trajectories. {\it Phys. Rev. A} {\bf96}, 053807 (2017).
\bibitem{Jor16}    
Jordan, A. N., Chantasri, A., Rouchon, P. \& Huard, B. Anatomy of fluorescence: quantum trajectory statistics from continuously measuring spontaneous emission.
{\it Quantum Stud.: Math. Found.} {\bf 3,} 237-263 (2016).
\bibitem{Jor13}
Chantasri, A., Dressel, J. \& Jordan, A. N.
Action principle for continuous quantum measurement.
{\it Phys. Rev. A} {\bf 88}, 042110 (2013).
\bibitem{Jor15}
Chantasri, A. \& Jordan, A. N. Stochastic path-integral formalism for continuous quantum measurement.
{\it Phys. Rev. A} {\bf 92,} 032125 (2015).

\end{references}
\end{document}